# Weapons Radiochemistry: Trinity and Beyond

Susan K. Hanson and Warren J. Oldham

Nuclear and Radiochemistry Group
Los Alamos National Laboratory
Los Alamos, NM 87545

**Abstract**: On July 16, 1945, the Trinity nuclear test exploded in the desert near Alamogordo, NM. A variety of new diagnostic experiments were fielded in an effort to understand the detailed performance of the nuclear device. This article describes a series of radiochemical experiments that were designed to measure the efficiency and neutron fluence of the test. These experiments, and the scientists who led them, laid the foundation of weapons radiochemistry for decades to come.

## Introduction

In March 1944, plans were initiated for the Trinity nuclear test. Because of major challenges encountered in calculating and predicting the performance of the plutonium implosion weapon, there was a growing consensus among scientists for the need to perform a test.[1–7]

A year later, in March 1945, initial concerns that the bomb would fizzle had evolved into more extensive technical discussions of which experiments would be fielded to understand performance and how to best coordinate the various diagnostic efforts. But what parameters were the most important to understand? A technical committee led by Harvard physicist Kenneth Bainbridge met weekly and considered various proposals.[6,8] With the paramount importance of maintaining the schedule, top priorities included assessing the pressure of the blast wave, the efficiency of the reaction (and thus the overall yield), and the simultaneity of the detonators. Photographic analyses and various measures of the radiation released by the detonation would also be fielded.[7,8,11]

According to physicist Robert R. Wilson, "It was recognized from the beginning that the most promising measurement of the nuclear efficiency would come from the radiochemical determination, and hence the greatest effort was put into this experiment."[8] Herbert L. Anderson, a physicist who had worked with Fermi at the Chicago Met Lab, was put in charge of what was deemed the "Conversion" efforts.[8] These experiments aimed to determine the efficiency of the explosion through measurements of residual actinide and fission product isotopes in samples collected after the detonation.[12] The overall efficiency of the device ($\varepsilon$) was related to the fraction of the plutonium that fissioned (Eq. 1), and was defined as the total number of fissions measured in the sample ($f$) ratioed to the total amount of starting plutonium, calculated as the measured plutonium in the sample ($^{239}Pu$) plus the measured fissions $f$, Eq. 2.[13]

$$\varepsilon = \frac{fissions}{original\ ^{239}Pu\ nuclei} \qquad (1)$$

$$\varepsilon = \frac{f}{^{239}Pu + f} \qquad (2)$$

This article reviews the radiochemical approach and measurements used to assess the yield of Trinity. With this first nuclear test, the basic foundation of weapons radiochemistry was established. Measurements and calibration subsequently became more sophisticated with the passage of time and emergence of new technologies, but the basic concepts endure to this day.

## Part 1. Trinity

### 100-ton test

In 1945, a large-scale test of conventional high explosives (HE) was scheduled to evaluate and prepare the procedures and diagnostic experiments that would be used for the Trinity nuclear test.[7,14] One hundred tons of Composition B would be detonated in the desert near Alamogordo, NM, on May 7, 1945. Figure 1 shows preparations for the test. In addition to the conventional HE, the test was doped with 16 gallons of radioactive liquid contained in saran (plastic) tubing.[15] The dispersal of the radioactive products would be studied to better understand the spread of radioactive fallout that would result from the upcoming atomic detonation. Radiochemical measurement procedures designed to separate fission products from soil would also be tested.[16] It was anticipated that some separation or fractionation of the different fission products could occur because of the heat and turbulence of the explosion; understanding the deposition patterns and potential chemical fractionation were stated goals of the experiment.[16]

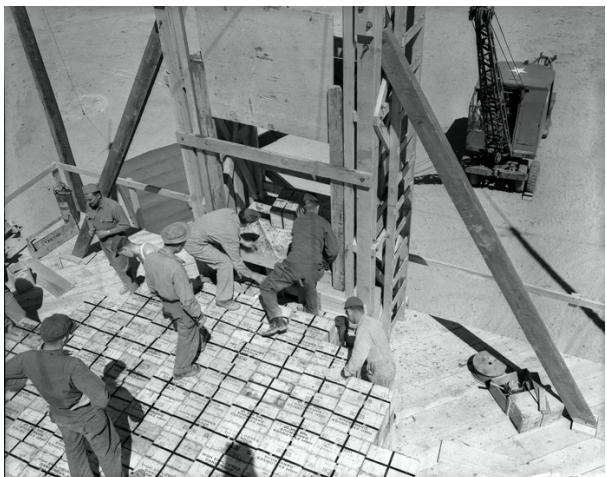

**Figure 1.** A photograph of the 100-ton test under construction.

To simulate the radioactivity that would be produced during the Trinity test, an irradiated fuel element or "slug" was obtained from the Hanford pile. The slug was irradiated for 116 days, incurring an estimated $1.8 \times 10^{14}$ fissions per second. The total radioactivity contained in the 100 tons of HE was thus approximately $1.8 \times 10^{21}$ total fissions, corresponding to 1000 curies of beta activity.[17] The slug was shipped from Washington to Alamogordo in a lead casket and moved with a crane truck into a special dissolving tank constructed with 3-foot concrete walls lined with 8 inches of lead. The slug was dissolved in concentrated nitric acid under remote operation. Once dissolution was complete, the pH of the solution was raised by addition of formic acid, and the solution was directly pumped into saran tubing that was interleaved through the HE pile. A higher pH was required by explosive experts, who were concerned that an accidental leak of the acidic radioactive mixture could cause a heating reaction and premature detonation of the high explosives.[17]

Immediately following the detonation, a team of scientists led by Anderson carried out surveys of the crater and collected samples of radioactive soils.[15] The team found that although the crater was smaller than they anticipated, so too was the total deposition of fission products. Geiger counter surveys of the area indicated that only about 1.5% of the total activity had deposited within a 150 foot radius. Most of the radioactivity was dispersed into the atmosphere, and the surface deposition was concentrated in a thin layer of fine particulate.[15]

Using newly developed methods, Sugarman's team performed radiochemical analyses on the samples collected from the crater of the 100-ton test. They measured a number of different fission products, including $^{89}$Sr, $^{91}$Y, $^{95}$Zr, $^{103}$Ru, $^{106}$Ru, $^{140}$Ba, and $^{141}$Ce.[18] Each fission product was chemically purified using a stable element carrier and a series of chemical separation steps to remove interferences from the soil and other fission product activities. The samples were counted using Geiger counters, and the measured activities for each fission product compared to the expected values (obtained from similar measurements performed on aliquots of the original solution of the dissolved slug). In general, the fission product measurement procedures were proven effective. Some loss of $^{95}$Zr was noted; it was not clear if this problem arose from an incomplete dissolution or from adsorption of the isotope on the container walls. Foreshadowing the Trinity test, some fractionation was documented between the ruthenium, barium, and lanthanide fission products.[18] This was ascribed to different chemical behavior of the elements at the high temperatures incurred during the explosion. Finally, procedures for purification and analysis of plutonium from soil samples were evaluated. The procedures were found to be effective, although the low plutonium levels in the samples from the 100-ton test (ca. 10 counts per minute/gram of soil) precluded any useful data assessment.[18]

In general, the 100-ton test was an impressive feat.[19] The experiment allowed scientists to assess basic assumptions of how a large amount of radioactivity would be dispersed in the explosion. In addition, the test served as a final validation of the radiochemical measurement methods that would be used to determine the efficiency of the Trinity test.

### Sample collection for Trinity

Collecting samples from the radioactive crater that would result from the first atomic bomb test was an entirely new challenge, and two primary sampling methods were planned.[7] Two Sherman M-4 tanks were lined with lead to protect the passengers from radiation. The first tank would drive directly into ground zero of the explosion and collect samples through a trap door on the bottom, using either a vacuum cleaner or a hollow pipe.[12] From the results of the 100-ton test, it was estimated that dose rates at the center of the crater would be 1000 R/hr.[15] The custom-built lead lining of the tank was found to shield the occupants of the tank from external radiation by a factor of about 40 times. The tank driver and passenger would receive further protection from airborne radioactivity by compressed air tanks that flowed into hood-type masks worn during the sampling.[12] In the event of a breakdown, the second tank could rescue the first tank and its passengers by towing it out of the radiation zone.[20]

The second sampling method consisted of using the other Sherman tank to fire rockets, each equipped with a sampling nose, into the center of the crater. The rockets were connected to a retractable cable and allowed sample collection from about 500 yards outside the crater. Each rocket was capable of trapping about 500 g of dirt in its collection body.[12] Figures 2–4 show photographs of a rocket, Sherman tank, and radiation survey meter, respectively, used to sample the Trinity crater.

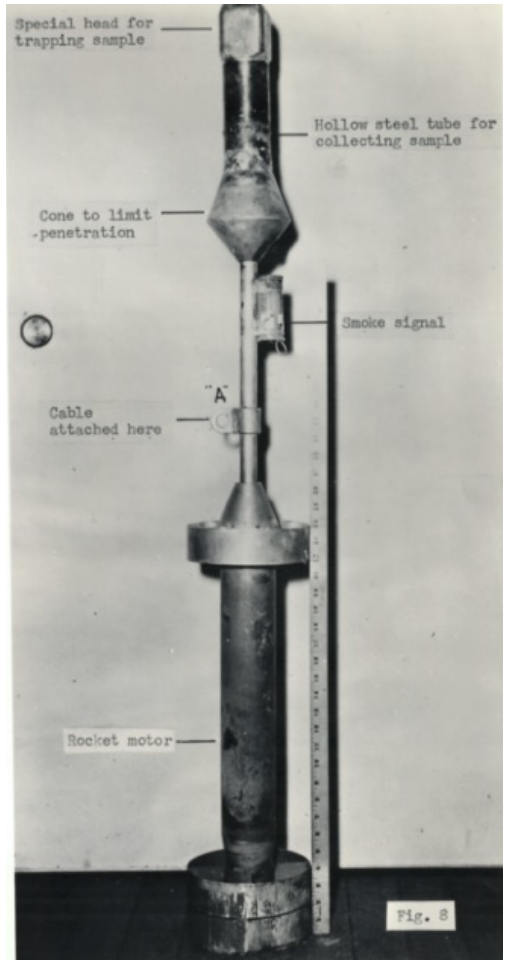

**Figure 2.** Photograph of one of the rockets used to sample the Trinity crater.

In addition to the tanks, sampling the ground by helicopter (from a height of ~200 m) was considered.[20] The main advantages of this method were that it would increase the distance from the radiation field, decrease the total time required for sampling, and decrease the exposure of the sampling personnel. Although the possibility of this technique was documented in a memo written by Anderson in April 1945, it was not further pursued, likely because of the unavailability of an appropriate helicopter in the timeframe of interest.[20]

On July 16, 1945, the Trinity test was fired, and 4 hours later the lead-lined tank made its first foray towards ground zero.[12] The tank reached about 100 yards from ground zero before having to turn back because of high radiation levels. According to Anderson, who led the radiochemical sampling effort, "During this run, it was discovered that the ground was covered with a fused layer of green stuff."[12] In a second run of the tank that occurred 12.5 hours after detonation, the tank reached the center of the crater. The radiation field inside the lead-lined tank was 25 R/hr, and the passengers received 5 R in a single 12 minute collection run.[12] Collection personnel worked staggered shifts to limit exposure to any one individual.

Rockets were fired from 1500 feet west of ground zero, and sampling proceeded smoothly, with multiple samples collected using this method. It was noted that the men operating the rockets received a much smaller radiation dose than the tank operators.[12] Although the tank engine broke down during the collection, five rockets were launched and four returned "excellent" samples.[12]

In addition to the ground samples, aerial sampling of the Trinity test was carried out. One airborne sample was collected about 30 miles north of the test site and contained an estimated $5 \times 10^{11}$ total fissions.[12] In addition, scientists J. Magee and A. Turkevitch realized that it might be possible to detect fission products produced by the Trinity nuclear test at a long range distance. According to LA-418, "The technique suggested was to filter large volumes of the upper atmosphere and count the activity in the filter."[21] As a result, a B-29 aircraft was equipped with a specially designed high-volume air filter unit in its forward bomb bay. On August 10, 1945, the aircraft was used to sample the atmosphere in British Columbia, where it was predicted that the Trinity cloud had diffused 24–30 days after the shot and after having circled the northern hemisphere.[21] The sampling mission was confounded by fission products from the Hiroshima detonation, on August 6, which were also expected to be present in the sampled air mass. Indeed, fission product activity attributable to Hiroshima was successfully detected on several of the filters. The scientists noted that the sampling method was "a practical means of detecting an atomic bomb explosion almost anywhere with proper meteorological conditions."[21] In addition, it was recommended that future aircraft sampling be performed closer to the point of detonation.

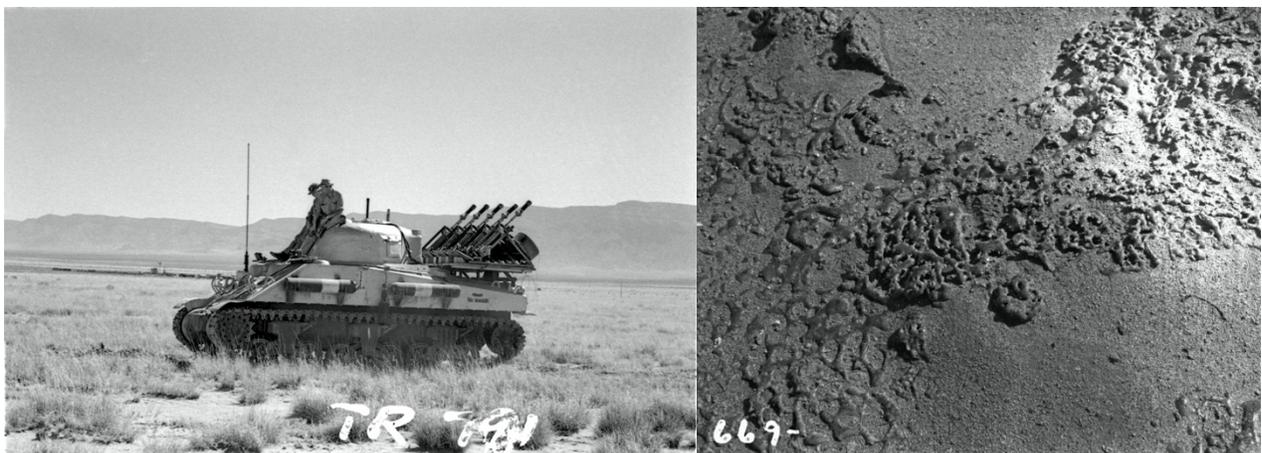

**Figure 3**. Left: Rocket sampling tank. Right: Fused, glassy debris observed on the ground after the Trinity detonation.

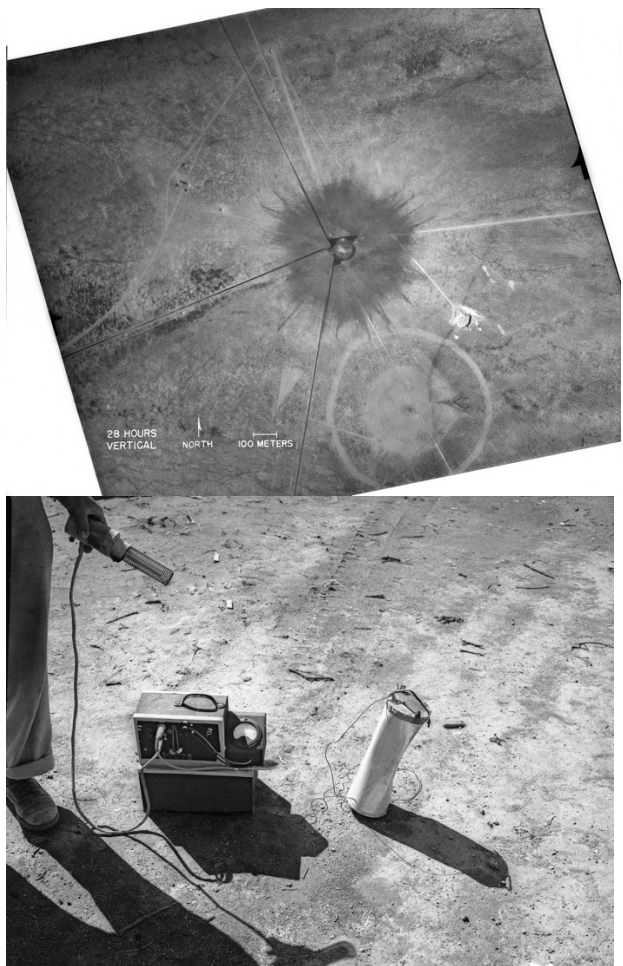

**Figure 4.** Top: Aerial view of the Trinity crater. The previous crater from the 100-ton test is visible in the lower right. Bottom: Radiation survey meter used at the Trinity test.

### Trinity fissions

The goal of the Trinity radiochemistry effort was to determine the efficiency, defined as the fraction of the nuclear device that fissioned. This would be assessed by measuring of the number of fissions relative to several fission products in samples collected from the crater, including [89]Sr, [97]Zr, [99]Mo, [140]Ba, [144]Ce, and [153]Sm.[12]

Measurement calibration was a major technical challenge. With considerable uncertainties existing in the known fission yields and half-lives of different isotopes, how could a measured activity of a particular isotope be related to the total number of fissions that occurred? Anderson's team performed a fission calibration experiment at the Omega Water Boiler reactor.[12] The high-power Water Boiler was an early critical assembly located in a canyon near Los Alamos and was composed of about 700 g of uranium enriched to 14.6% [235]U, dissolved in water, and shielded by a graphite and beryllium oxide tamper.[7] At the Water Boiler, a solution of plutonium dissolved in deuterated nitric acid ($DNO_3$) was irradiated.[12] A separate plutonium foil was mounted on the front of the vial and fission-counted.[22,23] The activity of different fission products measured in the solution was calibrated to the total fissions counted in the foil, where the relative amounts of plutonium in the solution and the foil were determined by comparison of their alpha counts. The thermal neutron spectrum in the experiment differed from the fast fission spectrum that would occur in the plutonium implosion bomb. However, the scientists noted that the calibration would be valid as long as there were not a significant change in fission yield with incident neutron energy and "it is expected that this condition would be fulfilled for fission products occurring at the peaks of the yield distribution curve."[12]

The samples collected immediately after the Trinity detonation ranged in concentration from about $5 \times 10^{10}$ to $2.5 \times 10^{13}$ fissions per gram; the highest concentrations were observed in the fused, glassy samples obtained closest to ground zero.[12] Good agreement was observed in the total numbers of fissions determined from [99]Mo, [144]Ce, and [97]Zr, although some issues were noted in the zirconium analysis caused by the presence of [239]Np and [95]Zr activities in the sample.[12] The activity of each purified fission product was measured in terms of counts per minute using Geiger-Müller tubes, and the activities were converted to fissions using the calibrations established in the Water Boiler experiments.[12]

Overall, the accurate determination of the total numbers of fissions in the Trinity test was a major technical accomplishment. The Los Alamos scientists developed creative approaches to obtain high-quality measurements, overcoming both limited knowledge of basic nuclear data and an entirely new experimental environment.

### Nathan Sugarman

One of the early radiochemists, Sugarman obtained his PhD from the University of Chicago in 1941. He moved to Los Alamos with his team in 1945, having spent the preceding years working as part of Enrico Fermi's group at the Chicago Met Lab.[1]

Although the exact origin of the plan to determine the efficiency of Trinity is somewhat unclear, both Anthony Turkevich and Richard Money credit Sugarman with the initial idea.[5,6]

According to Richard Money, *"Sugarman was on an official visit from Met Labs to Los Alamos. In a conversation with Oppenheimer, he convinced Oppenheimer . . . that by taking samples of the debris and doing radiochemical analyses and getting the radioisotope distribution that he could ascertain the efficiency. . . . Oppenheimer, I guess, told him 'Well, okay, you go ahead and do it, but don't get in the way of the physicists in their work.' Sure enough, Sugarman did take his measurements. As it turned out, Sugarman's measurements turned out to be the only meaningful measurements that were made at that time."*[5] Indeed, the 18,600 ton yield determined by Sugarman's team is remarkably close to the best modern value.

After the war, Sugarman returned to the University of Chicago as a chemistry professor. During his career, he had other significant scientific discoveries, including that of the isotope $^{85}Kr$ (10.7 year half-life).[6] He also held a reputation as an excellent teacher, striving to make all concepts as clear as possible.[10] He died in 1990, at the age of 73, after a distinguished scientific career.

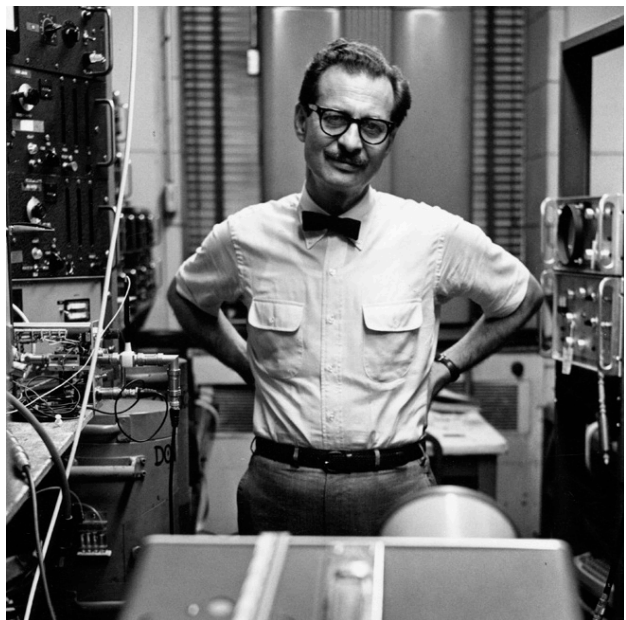

Photograph of Nathan Sugarman, ca. 1966. Credit: University of Chicago Photographic Archive, apf1-08002r, Special Collections Research Center, University of Chicago Library.

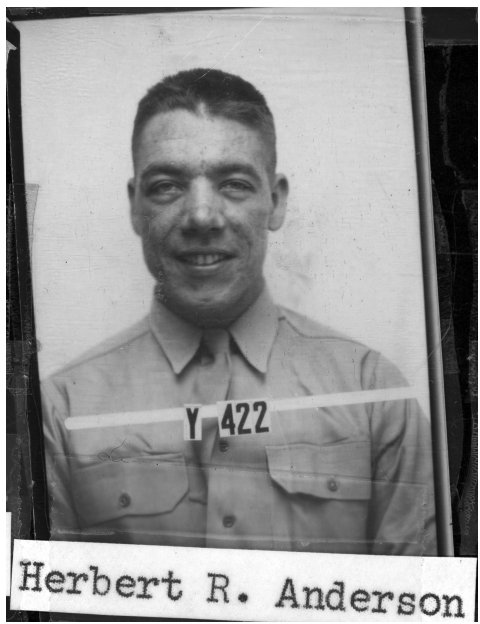

Photograph of Herbert Anderson, ca. 1944. Credit: Los Alamos National Laboratory Archives.

### Herbert Anderson

Herbert Anderson received his PhD in physics in 1940 from Columbia University, where he worked at the cyclotron with Professor John Dunning. He met Enrico Fermi at Columbia and moved to the Chicago Met Lab in 1942 as part of Fermi's group.[2] He worked briefly at DuPont on the Hanford effort and then returned to the Met Lab.[4] Anderson played a key role in the development of Chicago Pile 1 (CP-1), the world's first self-sustaining nuclear chain reaction located under the bleachers of Stagg Field.[2] Anderson moved to Los Alamos in 1944 and worked initially on the Omega Water Boiler critical assembly. Together with Nathan Sugarman, he led the effort to perform radiochemical analysis of Trinity and pioneered major concepts of weapons radiochemistry.

In a 1986 interview, Anderson described his experience at Trinity:[4]

**Anderson**: *What we did was, we collected out of the dirt . . . we picked up samples and analyzed the samples for plutonium and for the fission products. And by measuring the fission products and measuring the plutonium, you could tell what fraction of the plutonium actually underwent fission.*

**Interviewer**: *Did you arrive at the yield that way?*

**Anderson**: *Yeah, I produced the only really reliable figure. Everybody was surprised. It was much more powerful than anybody had expected.*[4]

After the war, Anderson returned to the University of Chicago as a professor, where he later became the director of the Enrico Fermi Institute (for nuclear physics).[2] Anderson returned to Los Alamos as a fellow in 1978. He died in 1988 from complications of chronic beryllium disease.

### Trinity actinides

For the efficiency determination, it was critical to relate the total fissions measured in a sample to the actinide fuel. The measurement effort in the Trinity test focused on residual plutonium. Plutonium measurements were performed by alpha counting using a Frisch grid ionization chamber filled with argon.[12] Some solutions of the nuclear debris from the Trinity test were evaporated directly onto a 5-inch-diameter platinum foil and alpha-counted; whereas other samples were chemically purified to remove up to 2 g of dirt prior to deposition. Two new purification procedures were developed to purify the soil: a ferric iodate coprecipitation and a cupferron extraction.[12] At the end of either purification, the plutonium was isolated as a $LaF_3$ precipitation and slurried onto a platinum disk for counting. The Frisch grid detector provided a gross assay of the total alpha emissions in the samples.

In an effort to quantify the amount of $^{238}Pu$ in the sample, the scientists performed a "range analysis," which allowed for some resolution of the alpha emissions on the basis of energy.[12] The results obtained from the Trinity samples were compared to a pure $^{238}Pu$ sample, and it was estimated that the samples contained approximately 2% $^{238}Pu$ (Figure 5).[12]

Anderson and coworkers also recognized that $^{240}Pu$ could be formed during the nuclear detonation.[12] Although the exact isotopic composition of the plutonium used for Trinity was not known, a fission counting experiment was undertaken to assess the $^{240}Pu$ content of the plutonium after the detonation. Samples of plutonium from the Trinity test and from the original ingoing material were fission-counted in the thermal column of the Argonne CP-3 pile.[12] In this thermal energy region, it was assumed that $^{240}Pu$ underwent minimal fissions relative to $^{239}Pu$. On the basis of the comparison of the fission and alpha counts of the two materials, it was estimated that the $^{240}Pu$ content of the postdetonation debris was about 6% of the total alpha activity, corresponding to a $^{240}Pu/^{239}Pu$ ratio of about 0.018.[12]

In addition to implementing new methods to measure changes in the isotopic composition of plutonium incurred during the detonation, the Manhattan Project scientists attempted to measure short-lived actinide species like $^{239}Np$ and $^{237}U$, formed principally through neutron activation reactions on $^{238}U$.[12] A chemical separation was performed to isolate $^{239}Np$ from the debris, but a lack of good methods to yield the sample hindered interpretation. Another strategy to measure $^{239}Np$ involved measuring the plutonium activity early on, then allowing the $^{239}Np$ in the sample to decay entirely to $^{239}Pu$, repeating the plutonium measurement, and comparing the two results. Although neither technique proved entirely satisfactory, it is significant that Anderson's team recognized the potential diagnostic power of this analyte in the first nuclear test. The $^{237}U$ isotope was also measured by counting in the days following the test.[12]

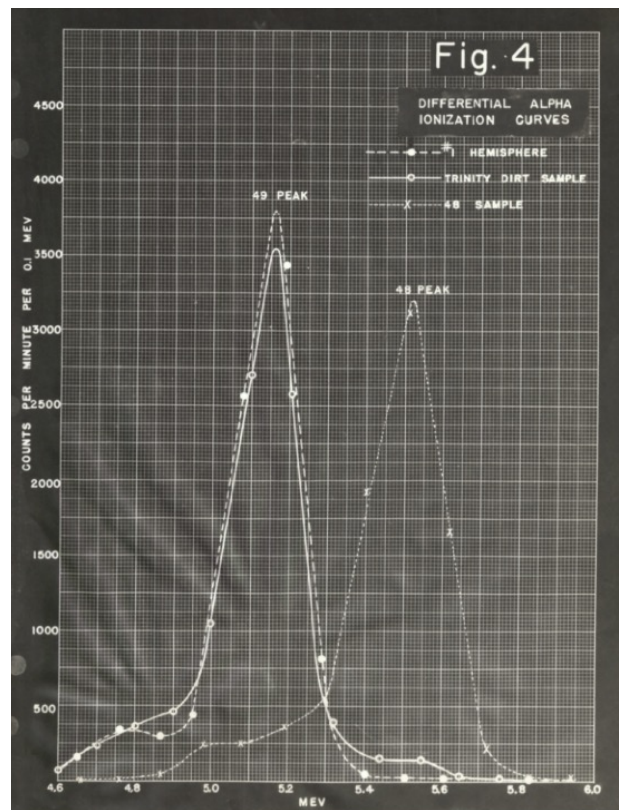

**Figure 5.** Original alpha spectrometry data from Anderson and Sugarman's 1945 report. The dashed line marked "hemisphere" is the alpha spectrum of the ingoing plutonium, the solid line is a Trinity debris sample, and the dotted line is a $^{238}Pu$ sample. The shoulder on the Trinity dirt sample under the 48 ($^{238}Pu$) peak is $^{238}Pu$ formed during the detonation.

### Monitors of neutron fluence and spectrum in Trinity—early appearance of detectors

In an effort to better understand the neutron fluence and spectrum, two additional radiochemical experiments were fielded during the Trinity nuclear test. One such experiment involved the placement of cadmium-coated gold foils fastened to iron pipes at 100-meter intervals from ground zero.[24,25] The induced radioactivity in the gold would be measured and used to calculate the total number of neutrons per square centimeter.[25] The gold foil experiments were led by Ernest Klema (Figure 6 and sidebar), who calibrated the activity measurements against a known Ra-Be neutron source.[25]

After the detonation, the lead-lined tank was sent in search of the gold foils. At least one of the gold foils was destroyed and those originally located at 400 m were blown far from their original placement.[24] The activity of the foils was plotted as a function of distance and decreased rapidly as the distance from ground zero increased.[24]

A second type of experiment led by Klema involved the placement of sulfur detectors. Sulfur was known to undergo a neutron capture reaction at a threshold energy of ~2.9 MeV and would provide information about fast neutrons (having energies greater than 3 MeV) in the detonation. The $^{32}S(n,p)^{32}P$ reaction cross sections had been studied in a series of experiments at the Wisconsin Van de Graaf accelerator located in the W building of the technical area, providing a calibration for the experiments.[9]

The detectors fielded in the Trinity test consisted of a short segment of end-capped iron pipe filled with elemental sulfur.[26] A total of eight detectors were fielded, with four secured 100 meters from the device and four secured 200 meters from the device. Only two of the detectors that were originally placed at 200 meters were successfully located after the test; these were returned to Los Alamos for analysis.[26] The measured $^{32}P$ activity was used to infer the total number of neutrons greater than 3 MeV that were detected on a sphere (radius = 200 meters) from the Trinity device.[26]

Although neither experiment provided decisive diagnostic information on Trinity, the placement of the gold and sulfur foils was a conceptual advance and established the potential utility of using radiochemical detectors to understand the neutron spectrum and fluence at various locations. The concepts established in these initial experiments would endure throughout subsequent nuclear weapons testing and development programs.

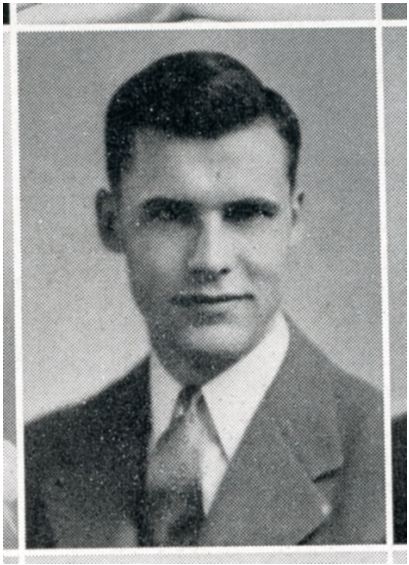

**Figure 6.** Photograph of Ernest Klema, 1941. The photo is from his senior year of college at Kansas University. Credit Jayhawkers Yearbook 1941, Kansas University Archive.

### Ernest Klema

In 1943, Ernest Klema was a first-year graduate student in physics working with Professor Robert Wilson at the Princeton cyclotron. The entire Princeton group moved to Los Alamos, where Klema was assigned to work with the Wisconsin Van der Graaf team. He worked to measure fission cross sections with the so-called "long tank," a detector system custom built for a neutron energy-independent response.[3]

Klema pioneered a set of innovative radiochemical detector experiments that were fielded during the Trinity nuclear test. He recounted: "*Some of my most vivid recollections are of the events surrounding the testing of the bomb at Alamogordo. I had the closest experiment to the tower holding the bomb that did not get destroyed in the explosion; consequently, I had to go in the next day to pick up my detectors. I saw a number of most unusual sights on that trip, including the finding of a number of round glassy beads, mostly green, but a few of blue or red color. These were formed by the melting of the desert floor as it was sucked into the fireball.*"[3]

The success of Klema's experiments at Trinity was attributable to a simple and robust design concept. Many years later, Klema commented on the flexibility of the approach: "*Each of us doing an experiment there had a theoretical advisor. The one for my experiment was Vicky Weisskopf. He and I got together one morning, and he gave me the results of his calculations so I could design the experiment. I went back to my lab and was working the afternoon when he called, 'Klema, this is Vicky. I just found out I made an error of ten to six [sic, $10^6$] in my calculations.' The ridiculous thing is that the experiment was able to accommodate quite easily.*"[3]

After the war, Klema returned to the University of Illinois to complete his studies. When the department chair did not allow his prior work on the cross section of the $^{32}S(n,p)^{32}P$ reaction to count towards his degree,[9] he transferred to Rice University, where he was awarded his PhD after more work building a large proton spectrometer.[3] He subsequently moved to the physics division of Oak Ridge National Laboratory, where he worked for a number of years in the 1950s, until becoming an assistant professor at the University of Michigan. During his second year at Michigan, he was awarded a tenured position at Northwestern University.[3] Klema continued to teach nuclear engineering and collaborate with Argonne National Laboratory in the area of nuclear reactions until 1968, when he moved to Tufts University. He died in 2008 after a prolific career.[3]

### *Trinity conclusions—modern perspective*

Seventy-five years later, one thing is clear about the radiochemical studies of Trinity: the measurements and assessment were remarkably accurate, particularly given the measurement tools and equipment of the time. Anderson and Sugarman's yield of 18,600 tons formed the basis of the official numbers published by Los Alamos and the Atomic Energy Commission/Department of Energy for the following decades. Very recent work by Selby et al. provides today's best assessment of 24.8 ± 2 kilotons.[27]

The Manhattan Project scientists also understood the limitations of their methods. They were clearly aware of potential problems arising from differences in chemical behavior of the different fission products and actinides in the fireball (chemical fractionation). To compensate, they focused their efficiency determination on fission product isotopes that lacked volatile gaseous precursors, and they compared the results of several different isotopes to recognize potential anomalies.

Anderson and Sugarman anticipated that changes in the plutonium isotopic composition would occur during detonation and characterized the plutonium with the best methods available. They estimated that the post-detonation samples contained approximately 6% $^{240}Pu$ and 2% $^{238}Pu$ (in terms of the total sample alpha activity). Modern analyses of fourteen samples performed in our laboratory reveal $^{240}Pu/^{239}Pu$ atom ratios that range from 0.02425(6) to 0.02508(6) with an average value of 0.0246(3). Some variability in the ratio occurs because of decay of $^{239}Np$ formed during neutron capture reactions on $^{238}U$ during the detonation. The average measured $^{238}Pu/^{239}Pu$ atom ratio, decay-corrected to July 16, 1945, is $2.75(4) \times 10^{-4}$. This corresponds to an activity ratio of 0.0693(1). The modern results are compared to those obtained by Anderson and Sugarman's team in Table 1.

In addition to fairly accurately determining the efficiency of Trinity, the Manhattan Project scientists predicted what would be important in the future. Sugarman's team attempted measurements of the short-lived actinide species $^{237}U$ and $^{239}Np$, both of which would be fundamental to weapons radiochemical diagnostics in the future. As a graduate student, Ernest Klema conceived and fielded a new type of experiment to capture neutron spectral and fluence information. These radiochemical detectors would be loaded in many experiments in the years to come. Together with the actinide and fission product measurements, they would form the backbone of weapons radiochemistry and diagnostics.

## Part 2. Moving Beyond

### *Sampling after Trinity*

After Trinity, aerial sampling was immediately implemented for nuclear weapon tests. The 1946 Operation Crossroads involved the detonation of two nuclear weapons at Bikini Atoll. Drones flown at different altitudes collected nuclear debris samples on filter papers, and the papers were analyzed at both a forward-deployed laboratory in the Pacific at Kwajalein and at Los Alamos. A similar air platform sampling approach was routinely used for subsequent atmospheric tests, although the drones were gradually abandoned in favor of manned flights equipped with filter units. These flights were typically flown directly into the debris cloud within hours of the detonation and were capable of collecting up to $10^{16}$ fissions per paper.

In 1948, ground sampling was also carried out as part of Operation Sandstone, using tanks that were remotely controlled via helicopter. The goals of the ground sampling effort were to serve as a backup to the aerial sampling and to collect a large amount of plutonium for isotopic analysis. In subsequent years, ground sampling was largely discontinued as aircraft sampling proved to be the more effective and safer method.

**Table 1.** Plutonium activity and atom ratios reported by Anderson and Sugarman and modern measured values. The modern values are the average measured values (and standard deviations) of fourteen different samples. All modern values are decay corrected to the reference date of 7/16/1945.

| Analysis | $^{238}Pu$ percent of total activity | $^{240}Pu$ percent of total activity | $^{238}Pu/^{239}Pu$ atom ratio | $^{240}Pu/^{239}Pu$ atom ratio | $^{238}Pu/^{239+240}Pu$ activity ratio |
|---|---|---|---|---|---|
| Anderson and Sugarman | 2 | 6 | $7.9 \times 10^{-5}$ | 0.018 | 0.020 |
| Modern LANL | 6.48(10) | 7.76(12) | $2.75(4) \times 10^{-4}$ | 0.0246(2) | 0.0693(1) |

Significant changes in radiochemical sampling occurred as nuclear tests moved underground in the years leading up to the 1963 Limited Test Ban Treaty. A variety of methods to sample underground experiments were attempted during the transition, including specially designed access ports, cables, U-traps, and "Hy-vac" vacuum apparatus. Ultimately, the most successful method proved to be an angled drill-back into the molten cavity formed during the test. The "Hunt sidewall sampler," a modified mining tool, was used to collect rock and fused glassy samples at various depths. Nuclear weapons tests were routinely sampled via drill-back until the start of the current testing moratorium in 1992.

### Fissions after Trinity

In the years following Trinity, assessments of nuclear efficiency and yield continued to rely on determination of the ratio of the total number of fissions in a sample to the residual actinide fuel. Considerable effort over decades focused on establishing and maintaining accurate calibrations of the fission measurements. Some of the fission products first measured in nuclear debris by Anderson and Sugarman ($^{89}$Sr, $^{97}$Zr, $^{99}$Mo, $^{140}$Ba, $^{144}$Ce, and $^{153}$Sm) remained important analytes for weapons radiochemistry. The analyte list was gradually expanded to include additional fission product isotopes, including $^{91}$Y, $^{95}$Zr, $^{111}$Ag, $^{115}$Cd, $^{147}$Nd, $^{156}$Eu, and $^{161}$Tb.

Fission product calibrations became more sophisticated as time elapsed. In the late 1940s, the concept of $k$-factors was developed. A $k$-factor was an experimentally derived value that related the observed activity of a particular fission product (in units of cpm) to the total number of fissions (measured with a fission counting chamber). This conceptual advance allowed for the total number of fissions in a sample to be accurately determined without reliance on the counter efficiency or fission yield, both of which were admittedly "subject to considerable uncertainty." As counting technology changed, the $k$-factors were repeatedly determined and validated in a series of irradiation experiments that occurred in both thermal reactors and critical assemblies. By the late 1970s, a sophisticated interlaboratory calibration campaign involving the National Institute of Standards and Technology (NIST) was carried out at the Big Ten critical assembly and LANL Omega West Reactor and used to establish both a lasting $^{99}$Mo $K$-factor at Los Alamos and a standardized "fission basis" across multiple laboratories.[28]

Fission product measurements remained a central focus, but the measurement technologies evolved over time. The Geiger-Müller tubes originally used to count samples and perform gross assays of radioactivity during the Manhattan Project era were ultimately replaced by more stable gas-proportional counters for measurements of beta decay emissions in purified samples of standardized mass and mount form. One of the first of such counters, described in a 1951 report by John Larkins,[29] is still functional and maintained in the Nuclear and Radiochemistry Group's count room today. Detection and quantification of gamma rays, initially performed with NaI scintillator detectors, transitioned to GeLi drift counters in the late 1960s.[13] The GeLi counters, precursors to modern high-purity germanium gamma-ray spectrometry, provided a revolutionary advance in energy resolution and a significant expansion in measurement capability.

For ease of comparison between experiments and with other laboratories, the concept of R-values was established. R-values are fission product measurement ratios designed to be both laboratory and counter independent. The earliest record of this approach to data analysis is documented in a 1949 report by R. Spence.[30] This report describes an experiment in the 14-MeV neutron energy regime, where R-values were defined as a double ratio of the activity of an isotope of interest to a reference isotope ($^{97}$Zr) in the experimental spectrum (14 MeV) related to the same activity ratio measured in a thermal spectrum.

Ultimately, the standard practice was to define an R-value as the ratio of an isotope of interest relative to a reference isotope (typically $^{99}$Mo or $^{147}$Nd) in an unknown experiment relative to the same ratio in a standardized $^{235}$U thermal experiment (Eq. 3). When the activities are measured using a consistent set of counters, the dependence on the counter efficiency and isotope-specific nuclear data that is present in both the numerator and denominator cancels out.

$$R = \frac{\left( A_X / A_{ref} \right)_{\text{experiment}}}{\left( A_X / A_{ref} \right)_{\text{U-235 thermal}}} \tag{3}$$

$A_X$ = activity of an isotope of interest
$A_{ref}$ = activity of a reference isotope, often $^{99}$Mo or $^{147}$Nd

Spence was one of the first Los Alamos radiochemistry group leaders and played an integral role in establishing these fundamental concepts and the calibrations of fission product measurements.

### Larkins' Counter #6

During the Manhattan Project, Geiger-Müller tubes were used for radiochemical measurements. Although these radiation detectors were inexpensive and readily available, the tubes suffered from limited long-term stability and a lack of energy resolution. By the mid-1950s, this measurement technique was largely surpassed by gas-proportional counters. One of the first counters of this type was counter #6, for which complete design specifications are described in the technical report LA-1238. The counter body was machined from dural (a hardened aluminum alloy), with a thin window and gaskets crafted to prevent gas leaking. The design intent was to achieve greater reproducibility between units and greater stability over time.

The original counter #6 is still maintained in the Nuclear and Radiochemistry Group's count room today. Although various parts have been refurbished over the years, the counter still functions after nearly 60 years, a testament to the success achieved with the original design concepts.

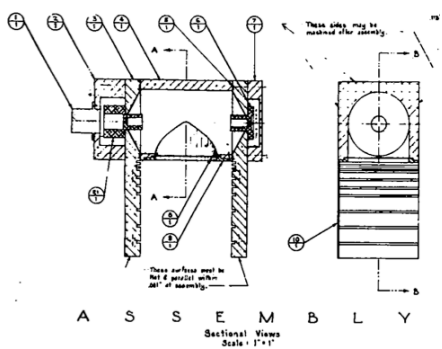
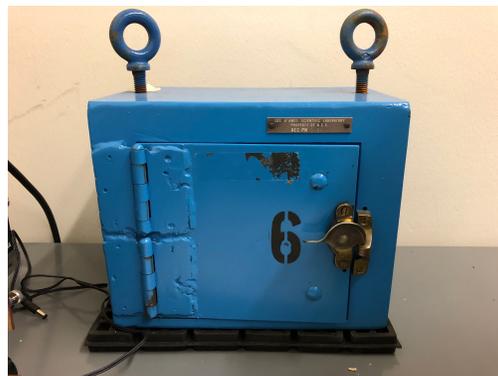

Left: Original design concept of the gas-proportional beta counter. Right: photo of counter #6, December 2020.

### Actinides after Trinity

Perhaps the most significant measurement evolution over time involved the actinides. For many years after the Trinity nuclear test, plutonium isotopic composition was determined using a combination of gross alpha counting, "differential ionization" (a form of energy-resolved alpha counting), and fission counting. In 1948, the first measurements of $^{235}U$ in nuclear debris samples were performed by fission counting, although it was acknowledged that "we do not [yet] know how to get $^{234}U$ and $^{236}U$."

Within just a few years, mass spectrometry emerged as a powerful technique to elucidate the isotopic composition of both uranium and plutonium. Actinide mass spectrometry measurements were originally performed as a specialized analysis at Argonne National Laboratory; the measurements moved to Los Alamos as the technique become more mature and widespread in the late 1960s.[13] More recent advances in mass spectrometry, including progress in Thermal Ionization Mass Spectrometry (TIMS) and the development of Inductively Coupled Plasma Mass Spectrometry (ICP-MS) have led to enhancements in the precision, sensitivity, and speed of analysis. At the same time, the sensitivity and resolution of alpha counting also improved greatly, particularly as a result of silicon semiconductor detectors becoming widely available. In the modern radioanalytical laboratory, both mass spectrometry and alpha spectrometry are routinely used to measure the concentrations and isotopic compositions of actinide elements.

### Detectors after Trinity

The ideas first demonstrated by Klema, of measuring induced activities in particular elements to ascertain neutron spectral and fluence information, transformed into a robust tool for nuclear physics and weapons experimental diagnostics. A foil or small amount of a particular element would be placed in a location of interest in an experiment and the induced activations used as a measure of the neutron spectrum and fluence at that site. Such approaches were applied in a wide range of irradiation experiments in nuclear reactors and critical assemblies in the decades that followed. By the 1950s, detectors were routinely fielded as part of nuclear weapons tests. Advanced analytical procedures were developed to separate and isolate detector elements from high activity nuclear debris. The Los Alamos capability grew when an isotope separator was purchased in 1964, expanding the suite of potential elements that could be used as detectors.[13] Research and development in the area continued until at least the mid-1980s, when the potential of new elements like bismuth and rubidium to serve as detectors was still being evaluated.[31,32]

## Other applications of radiochemistry—today's missions

In the years following Trinity, radiochemistry continued to grow as a variety of new applications were discovered. Research expanded into new frontiers of medical isotope production, nuclear power and waste mitigation strategies, nuclear chemistry, and environmental monitoring. At Los Alamos, today's Nuclear and Radiochemistry Group maintains expertise in radioanalytical chemistry and mass spectrometry and has ongoing missions in nuclear forensics, chemistry, and weapons diagnostics.

## Conclusions

The 1945 Trinity nuclear test launched a new scientific and political era. Groundbreaking contributions by Anderson, Sugarman, and Klema initiated the field of weapons radiochemistry. Looking back from today's perspective, it is remarkable not just how much they accomplished, but the extent to which their early studies laid the framework of what was to come over the course of the next 75 years.

## Acknowledgements

The authors wish to thank Roger Meade, Khal Spencer, Dana Labotka, Hugh Selby, and Chris Waidmann for helpful discussions.

This work was supported by the US Department of Energy through the Los Alamos National Laboratory. Los Alamos National Laboratory is operated by Triad National Security, LLC, for the National Nuclear Security Administration of the U.S. Department of Energy under Contract No. 89233218CNA000001.